\def\year{2019}\relax
\documentclass[letterpaper]{article} 
\usepackage{aaai19}  
\usepackage{times}  
\usepackage{helvet}  
\usepackage{courier}  
\usepackage{url}  
\usepackage{graphicx}  
\usepackage{multirow}
\usepackage{subfiles}
\usepackage{amssymb}
\usepackage{amsmath,bm}
\usepackage{color}

\newcommand{\qyl}[1]{{\color{black}#1}}

\def\eg{\emph{e.g}}

\begin{document}	
	%
	\title{Learning Bidirectional LSTM Networks for Synthesizing 3D Mesh Animation Sequences}
	\author{Yi-Ling Qiao\textsuperscript{1,2}, Lin Gao\textsuperscript{1}\thanks{Corresponding Author}, Yu-Kun Lai \textsuperscript{3}, \and Shihong Xia\textsuperscript{1}\\
		\textsuperscript{1}Beijing Key Laboratory of Mobile Computing and Pervasive Device, \\
		Institute of Computing Technology, Chinese Academy of Sciences \\
		\textsuperscript{2}School of Computer and Control Engineering, University of Chinese Academy of Sciences\\
		\textsuperscript{3}School of Computer Science \& Informatics, Cardiff University\\
		qiaoyiling15@mails.ucas.ac.cn, \{gaolin,  xsh\}@ict.ac.cn, LaiY4@cardiff.ac.uk
	}
	\maketitle
	\begin{abstract}
		In this paper, 
		we present a novel method for learning to synthesize 3D mesh animation sequences with long short-term memory (LSTM) blocks and mesh-based convolutional neural networks (CNNs).
		Synthesizing realistic 3D mesh animation sequences is a challenging and important task in computer animation. 
		To achieve this,  researchers have long been focusing on shape analysis to develop new interpolation and extrapolation techniques. However, such techniques have limited learning capabilities and therefore can produce unrealistic animation.
		Deep architectures that operate directly on mesh sequences remain unexplored, due to the following major barriers:  meshes with irregular triangles, sequences containing rich temporal information and flexible deformations. To address these, we utilize convolutional neural networks defined on triangular meshes along with a shape deformation representation to extract useful features,
		followed by LSTM cells that iteratively process the features. To allow completion of a missing mesh sequence from given endpoints,
		we propose a new weight-shared bidirectional structure.
		The bidirectional generation loss also helps mitigate error accumulation over iterations.
		Benefiting from all these technical advances, our approach outperforms existing methods in sequence prediction and completion both qualitatively and quantitatively. Moreover, this network can also generate follow-up frames conditioned on initial shapes and improve the accuracy as more bootstrap models are provided, which other works in the geometry processing domain cannot achieve.
	\end{abstract}
	\section{Introduction}\label{sec:introduction}
	Synthesizing high-quality 3D mesh sequences is of great significance in computer graphics and animation. In recent years, many techniques~\cite{Bogo2014,Dou2016,Stoll2010} have been developed to capture 3D shape animations, which are represented by sequences of triangular meshes with detailed geometry.
	Analyzing such animation sequences for synthesizing new realistic 3D mesh sequences is very useful in practice for the film and game industry.
	Although deep learning has achieved significant success in synthesizing a variety of media types, 
	directly synthesizing mesh animation sequences by deep learning methods remains unexplored. 
	In this paper, we  propose a novel long short-term memory (LSTM)~\cite{hochreiter1997long} architecture to learn from mesh sequences and perform sequence generation, prediction and completion.
	
	A major challenge to achieve this is to go beyond individual meshes and understand the \emph{temporal} relationships among them. Previous work on mesh data tries to perform clustering and shape analysis~\cite{huang2015analysis,sidi2011unsupervised} on the \emph{whole} datasets.
	However, none of them pay attention to temporal information, which is crucial for animation sequences.
	Thanks to the development of deep learning methods such as  the recurrent neural network (RNN) and its variants LSTM~\cite{hochreiter1997long} and gated recurrent unit (GRU)~\cite{cho2014learning}, one can more easily manipulate sequences. Based on RNNs,  impressive results have been achieved in tasks with regard to video, audio and text, \eg~movie prediction~\cite{mathieu2015deep,oh2015action}, music composition~\cite{lyu2015modelling}, text generation~\cite{vinyals2015show} and completion~\cite{melamud2016context2vec}.
	
	However, applying 
	deep learning methods to triangle meshes is not a trivial task due to their irregular topology and high dimensionality. Researchers often use fully connected networks in text or audio. Different from them, 3D shapes have spatial locality, which is suitable to work with convolutional neural networks (CNNs). However, unlike 2D images, shapes do not have regular topology.
	Recent effort has been made for lifting 2D CNN to 3D data~\cite{kalogerakis20173d}, including multi-view~\cite{su2015multi} or 3D voxel~\cite{riegler2017octnet,wu2016learning} representations.
	Alternatively, meshes can be treated as graphs, and based on this  a recent review~\cite{bronstein2017geometric} summarizes state-of-the-art deep learning methods in spectral and spatial domains. In order to reduce the number of parameters and extract intrinsic features, we utilize a CNN~\cite{duvenaud2015convolutional} defined on a shape deformation representation~\cite{gao2017sparse} that can effectively represent flexible and large-scale deformations.
	
	In summary, to analyze 3D mesh animation sequences, we propose a novel bidirectional LSTM architecture combined with mesh convolutions. The main contributions of this paper are:
	\begin{enumerate}
		\item We propose the first method to cope with mesh animation sequences, which allows generating sequences conditioned on given shapes, completing missing mesh sequences based on keyframes with realism and diversity and improving the generation of mesh sequences as more initial frames are provided. These capabilities significantly advance state-of-the-art techniques.
		
		\qyl{\item We design a share-weight bidirectional LSTM architecture that is able to boost performance and generate two sequences in opposite directions. Bidirectional generation also stabilizes training process and helps to complete a sequence in a more natural way.}
	\end{enumerate}
	
	\qyl{In the following, we first review relevant work, then presents our feature representation, network architecture, and loss functions. In Experiments section, we show extensive experimental results to justify our design and compare our work with previous work both qualitatively and quantitatively. Finally, we draw conclusions of our work.}

	\section{Related Work}\label{sec:relatedwork}
	
	\textbf{Sequence Generation with RNNs}.
	The recurrent neural network and its variants, such as  LSTM~\cite{hochreiter1997long} and GRU~\cite{cho2014learning}, have been widely used in dealing with sequential data, including text~\cite{bowman2015generating,mikolov2011extensions}, video~\cite{mathieu2015deep,oh2015action} and audio~\cite{chung2015recurrent,marchi2014multi}. \cite{srivastava2015unsupervised} learn representations of video by LSTM in an unsupervised manner. PredNet~\cite{lotter2016deep} learns to predict future frames by comparing errors between prediction and observation. \cite{yu2017seqgan} incorporate policy gradients with generative adversarial nets (GAN)~\cite{goodfellow2014generative} and LSTM to generate sequences. Attempts have also been made to predict video frames using CNNs~\cite{vondrick2016generating}. To avoid predicting videos directly in the high-dimensional pixel space, some work uses high-level abstraction such as human poses~\cite{walker2017pose,cai2017deep} to assist with generation.
	
	In the human motion area, researchers utilize RNNs to predict or generate realistic motion sequences. \cite{fragkiadaki2015recurrent} propose an encoder-recurrent-decoder (ERD) to learn spatial embeddings and temporal sequences of videos and motion capture. \cite{gregor2015draw} generate image sequences with a sequential variational auto-encoder, where two RNN chains are used to encode and decode the sampled sequences accordingly. However, such approaches that iteratively take the output as input to the next stage could cause error accumulation and make the sequence freeze or diverge. To address this problem, \cite{li2017auto} present Auto-Conditioned RNNs (acRNNs) whose inputs are previous output frames interleaved with ground truth. With ground truth frames at the beginning of a sequence, acRNN can also generate output sequences conditioned on given input sequences. \cite{martinez2017human} build a sequence-to-sequence architecture which is able to predict multiple actions, but they do not have spatial encoding modules. Using an encoder-decoder structure, \cite{butepage2017deep} extract feature representations of human motion for prediction and classification. \cite{cai2017deep} use GAN and LSTM to generate actions or complete sequences by optimizing the input vector of the GAN.

	\textbf{3D Shape Generation}.
	Generating 3D shapes is an important task in graphics and vision community. Its down-stream applications include shape prediction, reconstruction and sequence completion. Nevertheless, such tasks are 
	more challenging due to the high dimensionality and irregular connectivity of mesh data. Previous work mostly generates 3D shapes via interpolation or extrapolation in parameterized representations. 
	\cite{huber2017smooth} propose to interpolate shapes in a Riemannian shell space. Based on existing shapes, data-driven methods (e.g.~\cite{gao2017data}) can generate realistic samples. However, such traditional methods focusing on shape representations and shape analysis have limited learning capabilities.
	More recently, \cite{tan2017variational} propose to use Variational Autoencoders (VAEs) to map mesh models into a latent space and generate new models by decoding latent vectors. Locally deformed shapes can also be generated by a combination of deep learning and sparse regularization~\cite{tan2017mesh}.
	While these learning based methods can produce new shapes which are more diverse and realistic, the temporal information of mesh animation sequences is not fully explored.
	

	\section{Methodology}\label{sec:methodology}
	\begin{figure*}[t]
		\begin{center}
			\includegraphics[width=0.8\linewidth]{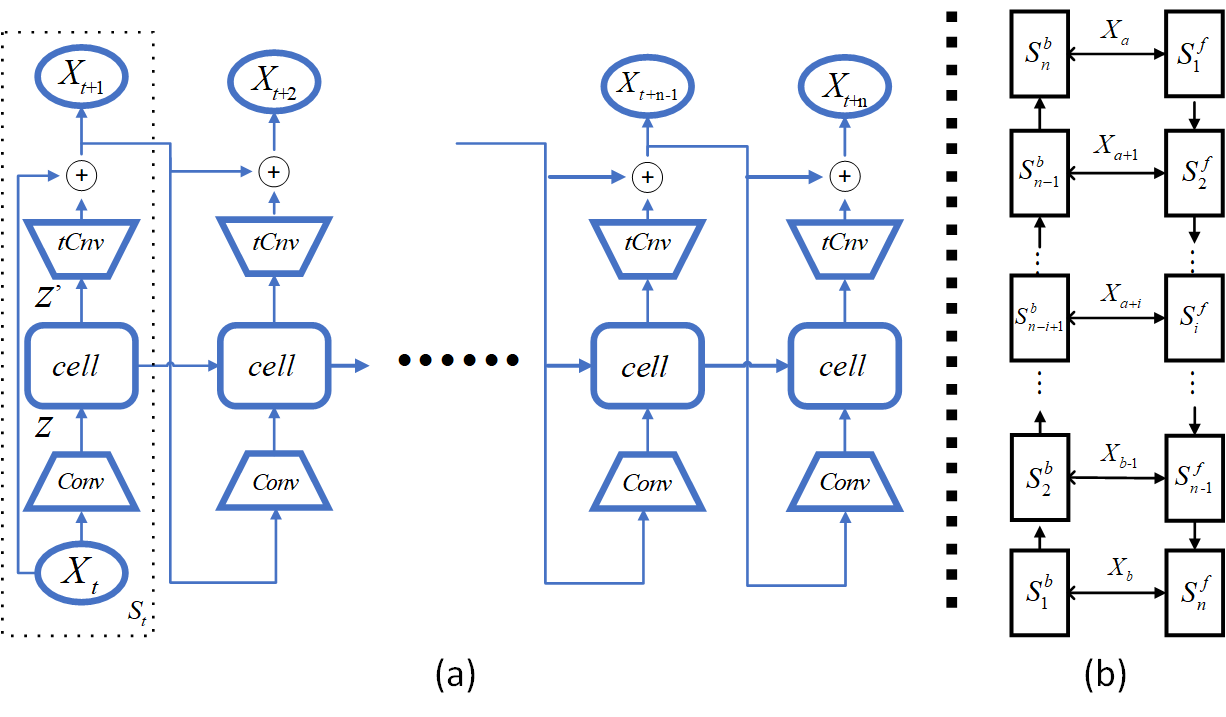}
		\end{center}
		\caption{Architecture of our network. (a) shows that our network is composed of LSTM module $cell$ and mesh convolution module $Conv,tCnv$. Take the network $S_t$ at time step $t$ as an example, the input to $Conv$ is the deformation representation $X_t$. The interface between $cell$ and $Conv$ is a fully connected layer, which outputs a low-dimensional vector $z$ into $cell$. $tCnv$, a stack of transpose convolution layers, mirrors $Conv$ and shares weights with it. The output of $tCnv$ is the feature change $\delta X_t$. $\delta X_t+X_t$ gives the predicted feature for time step $t+1$, which is fed into $S_{t+1}$ iteratively. (b) is our bidirectional LSTM. Both chains have the same architecture as in (a), and the only difference is their opposite direction. The forward chain takes the first model as input and the backward chain takes the last. They share weights and their predictions are constrained to match with each other. }
		\label{fig:networkstruct}
	\end{figure*}

	\subsection{Mesh Sequence Representation}\label{sec:representation}
	
	Mesh animation sequences are typically represented as a set of meshes with the same vertex connectivity and different vertex positions. Such meshes can be obtained by consistent remeshing or mesh deformation, and become very common nowadays due to the improved scanning and modeling techniques. These animated mesh sequences usually contain large-scale and complex deformations.
	
	\qyl{In this work, we represent shapes using a shape deformation representation~\cite{gao2017sparse}, a state-of-the-art representation which works well for large-scale deformation and suitable for deep learning methods.}
	
	
	Assume the mesh sequence dataset $M$ contains $n$ shapes and each mesh is denoted is as $m_t$ ($t=1, 2, \dots, n$). We denote $\bm{p}_{t,i}\in \mathbb{R}^3$ as the $i^{\rm th}$ vertex of the $t^{\rm th}$ model. $\bm{D}_{t,i}$ represents  the deformation gradient defined in each 1-ring vertex neighborhood, which is computed as
	
	\begin{equation}\label{equa:T}
	\underset{\bm{D_{t,i}}}{\arg\min}=\underset{j\in N_i}{\sum}
	c_{ij}\left\|(\bm{p}_{t,i}-\bm{p}_{t,j})-\bm{D_{t,i}}(\bm{p}_{1,i}-\bm{p}_{1,j})\right\|_{2}^2
	\end{equation}
	where $N_i$ is the 1-ring neighbors of the $i^{\rm th}$ vertex of the $t^{\rm th}$ shape, and $c_{ij}$ is the cotangent weight to avoid discretization bias~\cite{Levi}. 
	The deformation gradient matrix $\bm{D_{t,i}}$ is decomposed into rotation matrix $\bm{R_{t,i}}$ and scaling matrix $\bm{S_{t,i}}$: $\bm{D_{t,i}}=\bm{R_{t,i}}\bm{S_{t,i}}$. The difficulty for representing large-scale deformations is that the same rotation matrix $\bm{R_{t,i}}$ is mapped to 
	two rotation axes with opposite directions, and the associated rotation angle can include different number of cycles. 
	To solve this rotation ambiguity problem, a global integer programming based method~\cite{gao2017sparse} is applied to obtain as-consistent-as-possible assignment which outputs a feature vector $\bm{q_{t,i}}\in\mathbb{R}^9$. The mesh representation $X_t$ is eventually produced by linearly normalizing each dimension of $\bm{q_{t,i}}$ into $[-0.95,0.95]$~\cite{tan2017variational}.

	\subsection{Generative Model}\label{sec:model}
	The overall architecture of our approach is illustrated in Fig.~\ref{fig:networkstruct}. In this illustration, we denote LSTM cells as $cell$. $Conv$ refers to the mesh convolutional operations~\cite{duvenaud2015convolutional,tan2017mesh} and $tCnv$ represents transpose convolutions. For each convolutional filter, the output at a vertex is computed by a weighted sum of its 1-ring neighbors along with a bias:
	\begin{equation}\label{equa:conv}
	y_{i} = W_{1}x_i+W_{2}\frac{\sum_{j=1}^{d_i}x_{n_{ij}}}{d_i}+b
	\end{equation}
	where $x_i$ and $y_i$ are input and output at the $i^{\rm th}$ vertex, $W_1$, $W_2$ and $b$ are the filter's weights and bias, $d_{i}$ is the degree of the $i^{\rm th}$ vertex, and $n_{ij}$ is the $j^{\rm th}$ 1-ring neighbor of the $i^{\rm th}$ vertex. The interface between LSTM module and mesh convolution layers is a fully connected layer.
	
	Given the LSTM state $s_t$ and model $X_t$, we first describe how to generate the next model $X_{t+1}$.
	First we put $X_t$ into the mesh convolutional sub-network $Conv$, which outputs a low-dimensional latent vector $z=Conv(X)$. After that, $z$ is sent to LSTM cell $cell$ and the output is in the following form: $(\hat{z}, s_{t+1})=cell(z, s_t)$, where $s_{t+1}$ represents the updated state and $\hat{z}$ is the updated latent vector. $\hat{z}$ is then passed to transpose mesh convolution $tCnv$. Similar to many sequence generation algorithms, the output of $tCnv(\hat{z})=\delta X_{t}=X_{t+1}-X_{t}$ is defined as the difference between the next and current models, instead of $X_{t+1}$ to alleviate error accumulation. In the end, the generated model from $X_t$ is simply worked out as $X_{t+1}=X_{t}+\delta X_{t}$. Consecutive models are generated iteratively in the same way.  For simplicity, the whole process in one iteration is denoted as $(s_{t+1},X_{t+1})=G(s_t,X_t)$\\

	Fig.~\ref{fig:networkstruct}  illustrates the whole process of generating sequential data using our model. Suppose that we already have a set of models $S_{i,j}=\{X_{i}, X_{i+1},..., X_{j}\},(i\leq j)$. To extend the sequence, we would like to predict its  $n$ future models $S_{j+1,j+n|i,j}=\{X_{j+1}, X_{j+2},..., X_{j+n}|X_{i},X_{i+1},..., X_{j}\}$. Our method first puts the existing models into the network in their order, lets the LSTM cell update its state to $s_j$ from an initial state $s_0$. When it comes to the $j^{\rm th}$ model, the network outputs $X_{j+1}$, which is afterwards treated as the $j+1^{\rm st}$ input, and this process repeats for $n$ times, leading to the follow-up sequence $S_{j+1,j+n|i,j}$.
	
	\subsection{Bidirectional Generation}\label{sec:bigen}
	Sequence generation is a promising while challenging problem in various data forms like video, music and text, not only for the potentially tricky way to exploit temporal information but also about how to obtain enough training data. 
	When the data is scarce for a specific application, which is often the case for 3D model datasets, training can be problematic.
	
	However, unlike text, movie and audio, 3D model sequences can be more flexible. On the one hand, the order of 3D shape sequences is less strict, i.e., the inverse of a motion can also be reasonable. On the other hand, there are usually multiple plausible paths between two shapes. Based on those two observations, we propose a bidirectional generation constraint, which avoids restricting results to specific deformation paths, as shown in Fig.~\ref{fig:networkstruct}. From a 3D model dataset, we arbitrarily choose two models $X_{a}, X_{b}$ as endpoints of two inverse $n$-length sequences $S^{f}, S^{b}$ such that $S^{f}_1=S^{b}_n=X_{a},S^{f}_n=S^{b}_1=X_{b}$. Let $X_{a}, X_{b}$ have opposite initial states $s^{f}_0,s^{b}_0$, we expect them to generate similar models, satisfying $\forall 1\leq i\leq n,S^{f}_i \approx S^{b}_{n+1-i}$.
	
	\subsection{Loss Function}\label{sec:loss}
	In this paper, the loss function is composed of three terms as
	\begin{equation}\label{equa:lossall}
	L = L_{reconstruct}+\alpha_1 L_{bidirection}+\alpha_2 L_{reg}
	\end{equation}
	To illustrate this, let the ground truth models be $\{X_1,X_2,...,X_n\}$, which are expected to be the results of forward sequence $S^f$ and backward sequence $S^b$.  The reconstruction loss $L_{reconstruct}=\sum_{i=1}^{n}(||S^{f}_{i}-X_i||+||S^{b}_{i}-X_{n+1-i}||)$ forces both sequences to resemble samples from the dataset. Meanwhile, as described before, bidirectional sequence  $S^f,S^b$ share weights and have similar outputs, which is ensured by bidirectional loss $L_{bidirection}=\sum_{i=1}^{n}||S^{f}_i-S^{b}_{n+1-i}||$. Furthermore, $L_{reg}=L_{KL}+L_2$ contains KL divergence term and $L_2$ loss to regularize the network. The KL divergence between the low-dimensional vector and Gaussian distribution is computed so as to get a good mapping. Therefore, we have $L_{KL}=D_{KL}(q(z|X)|p(z))$, where $q(z|X)$ is the posterior distribution and $p(z)$ is the Gaussian prior distribution. In experiments, we set $\alpha_1=0.5,\alpha_2=0.1$.

	\section{Experiments}\label{sec:experiments}
	\begin{figure*}[t]
		\begin{center}
			\includegraphics[width=0.95\linewidth]{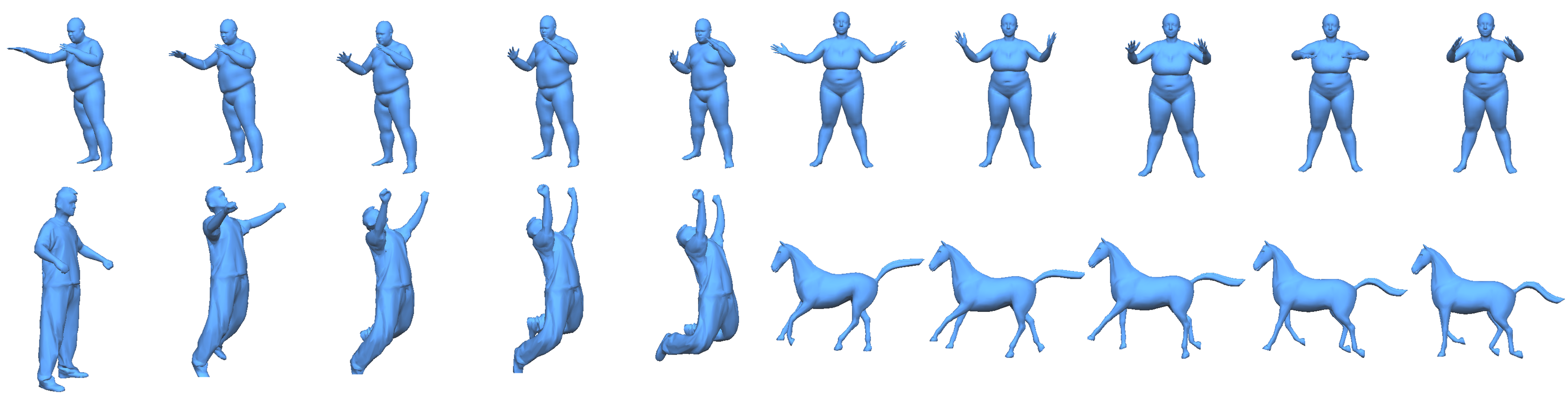}
		\end{center}
		\caption{Qualitative results of our method on Dyna~\cite{pons2015dyna}, handstand~\cite{vlasic2008articulated} and horse~\cite{sumner2004deformation}. We give one source model to the network and it generates the following four shapes. This is the first approach able to generate a whole sequence from only one mesh.}
		\label{fig:qual}
		
	\end{figure*}

	\begin{figure*}[t]
		\begin{center}
			\includegraphics[width=.95\linewidth]{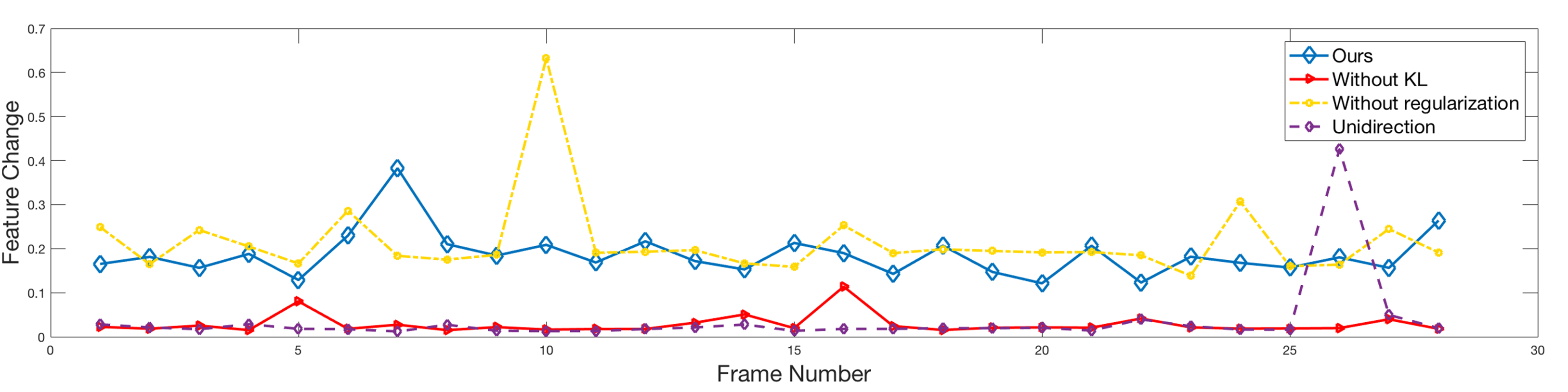}
		\end{center}
		\caption{Shape feature change between subsequent frames of different methods. This line graph depicts the amount of feature changes $mean(||X_{t}-X_{t+1}||/X_{t})$ between consecutive frames. The proposed method (blue) has stable and visible changes. Networks without KL loss or BD-training suffer from frozen sequences, and the one without $L_2$ regularization has significant jerk.}
		\label{fig:line}
	\end{figure*} 	
	
	\subsection{Framework Evaluation}\label{sec:evaluation}
	
	We now evaluate the effectiveness of different components in our framework.
	
\begin{figure*}[t]
		\begin{center}
			\includegraphics[width=0.9\linewidth]{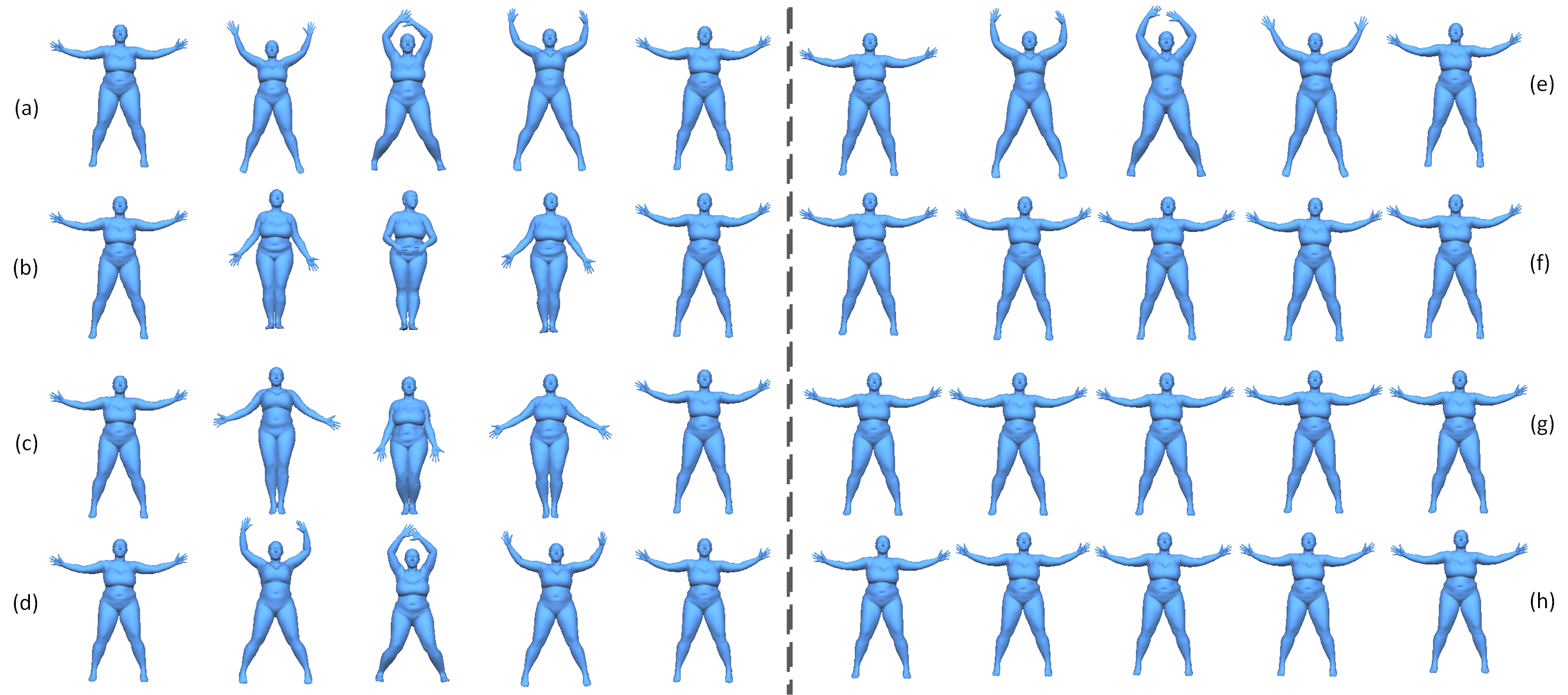}
		\end{center}
		\caption{Diversified sequence completion. We show the completion results produced by different methods. Source (first) and target (last) shapes are shared among all the sequences. (a) is the ground truth from f\_ 50004\_jumpingJacks dataset~\cite{pons2015dyna}; (b)(c)(d) use our BD-LSTM; (e) is the optimization+unidirectional baseline strategy described in
			\qyl{the paper}; 
			(f), (g) and (h) are interpolation results using \cite{tan2017variational}, \cite{gao2017data} and \cite{gao2017sparse} accordingly. We can see that (f)(g)(h) generate almost identical shapes because interpolation follows the shortest path between source and target. Compared to (e), our method can generate diverse, plausible results for users to choose.}
		\label{fig:completion}
	\end{figure*}

	\textbf{Bidirectional generation.} 
	\qyl{We propose a share-weight bidirectional LSTM (BD-LSTM) to better utilize temporal information and facilitate sequence completion.}
	Fig.~\ref{fig:completion} demonstrates that our BD-LSTM can produce results with better diversity. On the other hand, the $L_{bidirection}$ and $L_{KL}$ terms impose stronger constraints during training and consequently helps predict more accurate sequences. According to the numerical results in Tab.~\ref{tab:arch_test}, our method is more effective than existing methods~\cite{tan2017variational,gao2017sparse}. \qyl{Moreover, our method benefits from multiple initial frames, as well as the bidirectional constraint. In Tab.~\ref{tab:rebutaladd}, we show the results if we do not use our BD-LSTM or leave out $L_{KL}$ term.}

	\textbf{Loss terms.}
	Error accumulation is a common problem in sequence generation tasks~\cite{gregor2015draw,li2017auto,martinez2017human}. Generated meshes usually freeze because results tend to stay at an average shape, or even diverge to random results. To address this problem, we use three methods, 1) $L_{KL}$ divergence to regularize the internal distribution, 2) $L_2$ regularization loss to mitigate overfitting, and 3) bidirectional generation to impose an additional constraint. To justify those terms, we train models without one of the three. For unidirectional sequences, we only use one direction of LSTM. The line graphs in Fig.~\ref{fig:line} shows representation changes $mean(||X_{t}-X_{t+1}||/X_{t})$ between adjacent frames $X_{t}$ and $X_{t+1}$. The four networks are trained on Dyna~\cite{pons2015dyna} for 7000 iterations, and tested on 32 randomly chosen sequences. From the test one can see that without KL or BD-LSTM, the sequence tends to freeze. Meanwhile, $L_2$ regularization helps to reduce jerk. In Tab.

	\textbf{Initial frames.}
	Generating a sequence based on initial frames is an important application. In theory, the more bootstrap frames we have, the more knowledge we obtain about the sequence therefore we are supposed to make more accurate prediction. Previous mesh generation approaches, however, are based on interpolation/extrapolation, which can only use two of the existing models (endpoints). Our method can take advantage of all input frames by feeding them into the LSTM.  A previous human motion prediction method uses $u(=4)$ frames to start the recurrent network~\cite{li2017auto}.  We test $u=1$ and $u=3$ in Tab.~\ref{tab:arch_test} to show that more initial frames can reduce the distance between prediction and ground truth.

	\begin{table}[tb]
		\begin{center}
			\setlength{\abovecaptionskip}{2pt}
			\begin{tabular}{|c|c|c|c|}
				\hline
				Ours & Std. BD-LSTM & Unidir. LSTM & no $L_{KL}$ \\ \hline 
				\textbf{88} & 123 & 114 & 103 \\ \hline
			\end{tabular}
			\rightline{per-vertex position error($\times 10^{-4}$)}
			
			\caption{\qyl{Average vertex position errors on the Punching dataset~\cite{pons2015dyna} with different network architectures. We can see that our share-weight BD-LSTM outperforms standard BD-LSTM and unidirectional LSTM. Also we observe a decrease in accuracy if the $L_{KL}$ term is omitted.}}		
			\label{tab:rebutaladd}
		\end{center}
	\end{table}
	
	\begin{figure*}[t]
		\begin{center}
			\includegraphics[width=.95\linewidth]{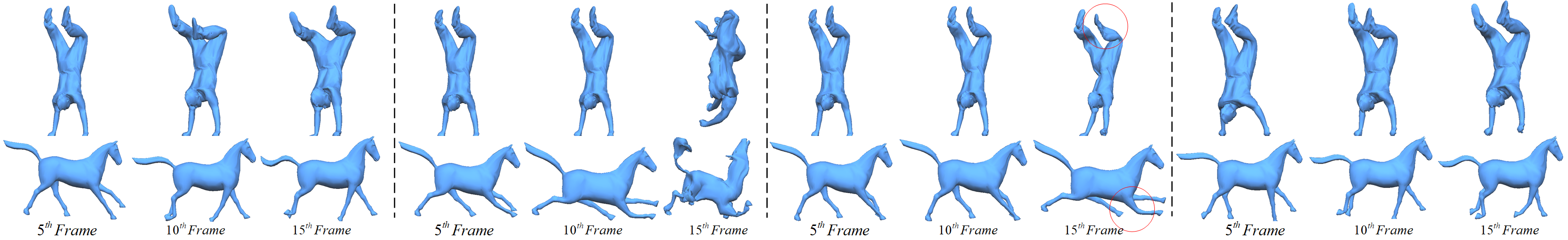}
		\end{center}
		\begin{tabular}{p{4.3cm}p{4.3cm}p{4.3cm}p{4cm}}
			(a) Ground Truth&(b) \cite{gao2017sparse}&(c) \cite{tan2017variational}&(d) Ours
		\end{tabular}
		\caption{Comparison with other work on sequence generation. In this experiment, two consecutive frames are sent into the network, and we aim to predict the future $5^{\rm th},10^{\rm th},15^{\rm th}$ shapes. (a) shows the ground truth of relevant shapes; (b) is obtained by using linear extrapolation on the feature~\cite{gao2017sparse}; (c) is extrapolation on a feature from deep learning~\cite{tan2017variational}; (d) is our result. We can see that extrapolation-based generation fails when predicting frames further away. (b) totally fails on the $15^{\rm th}$ frame. and (c) also produces abnormal deformation, as highlighted in the red circles. In contrast, our method  forms a natural cycle and avoids exceeding the limits (following the horse's stride).  }
		\label{fig:gene}
	\end{figure*} 	
	
	\begin{figure*}[t]
		\begin{center}
			\includegraphics[width=.9\linewidth]{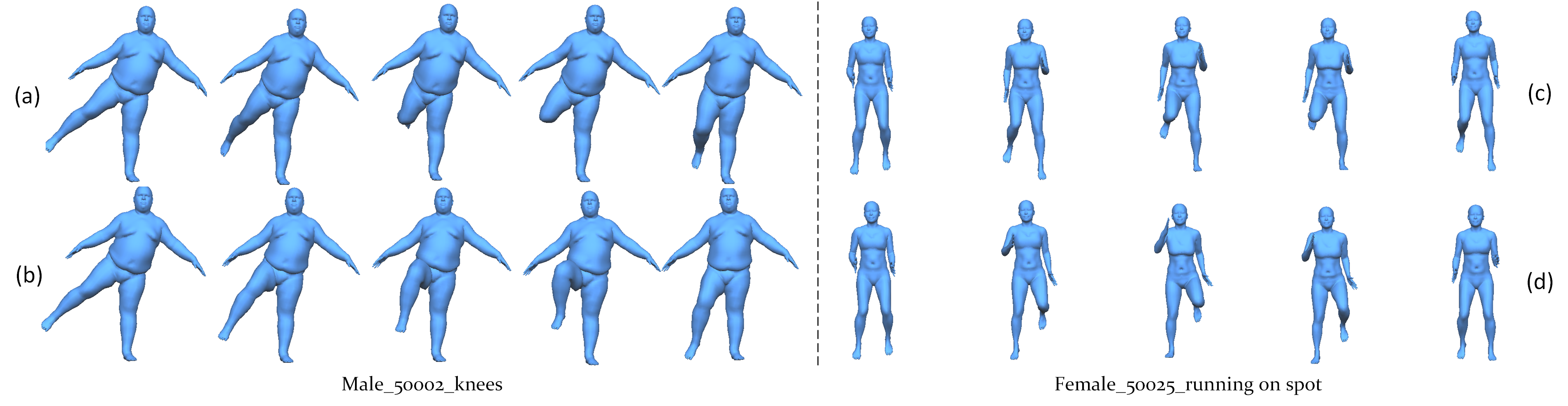}
		\end{center}
		\caption{Conditional generation. Trained with a mixture of different Dyna datasets, our network can output sequences conditioned on the first two input shapes. For examples, on the right of the figure, we feed two fit female shapes into the network. The second frame in (c) lifts her right leg while (d) lifts the left leg. Our model can perceive their differences and predict subsequent motion according to the condition. Similar results can be observed in the fat male example on the left.}
		\label{fig:conditional}
	\end{figure*} 	
	\begin{table*}[t]
		\begin{center}
			\setlength{\abovecaptionskip}{2pt}
			\begin{tabular}{|c|c|c|c|c|c|c|c|c|c|c|c|c|}
				\hline
				\multirow{2}{*}{Method} 	 & \multicolumn{3}{|c|}{Punching}  &   \multicolumn{3}{|c|}{ShakeArm}  &  \multicolumn{3}{|c|}{Handstand}   &  \multicolumn{3}{|c|}{Horse}  \\
				\cline{2-13}
				&5&10&15&5&10&15&5&10&15&5&10&15\\	 \hline \hline
				Ours+1 IF& 175& 156& 285& 335& 381& 319& 323  & 527& 516& 603& 869& 1328\\ \hline
				Ours+3 IF& 95& 84&  {\bf 107}& 301&  {\bf 226}&{\bf 290}& 212 &  {\bf 379}&  {\bf 428}&451&  {\bf 329}&  {\bf 671}\\ \hline \hline
				\cite{tan2017variational}& 132& 240& 457&  {\bf 291}& 433& 688&  {\bf 93}& 489& 797&  {\bf 286}& 713&1032\\ \hline
				\cite{gao2017sparse}& 294& 361& 413& 391& 472& 110& 487& 401&1589& 334& 1051& 1568\\ \hline
			\end{tabular}
			\rightline{per-vertex position error($\times 10^{-4}$)}
			
			\caption{Comparison of variants of our method and previous work on per-vertex position error (average distance between vertex positions of ground truth and prediction). In this experiment, we observe that more initial frames (IF) will improve the performance. Our method outperforms \cite{tan2017variational} and extrapolation+~\cite{gao2017sparse} since they suffer from error accumulation thus the accuracy degrades as the sequence moves on.  }		
			\label{tab:arch_test}
		\end{center}
	\end{table*}
	
	\subsection{Sequence Generation} \label{sec:generation}
	
	We now evaluate sequence generation capability of the proposed method. Starting from some initial frames, sequence generation predicts future frames. 
	
	\textbf{Generating sequences.}
	As far as we are aware, this is the first work to learn and generate arbitrarily long mesh sequences. Given two initial frames, people used to generate meshes through extrapolation~\cite{tan2017mesh}. However, simply extrapolating shapes fail to capture long-term temporal information, \eg~periodicity of the sequence. With the help of LSTM, our model can record history information and iterate to generate realistic mesh sequences in any length, even if the number of models in the dataset is limited. In the experiment, we feed first two mesh models to the LSTM and let it generate following frames. Qualitative and quantitative results are shown respectively In Fig.~\ref{fig:qual} and Tab.~\ref{tab:arch_test}. We compare our model with ground truth as well as previous extrapolation-based methods~\cite{tan2017variational,gao2017sparse}. Fig.~\ref{fig:gene} plots the predictions on the $5^{\rm th}$, $10^{\rm th}$ and $15^{\rm th}$ future frames. We can see that both extrapolation  methods fail on the $15^{\rm th}$ frame, because linearly extending the motion path eventually exceeds the plausible deformation space. In contrast, our method is aware of periodicity of the sequence, and able to return back once reaching the extreme point, producing natural motion cycles.

	\textbf{Conditional generation.} Another promising application of our method is to generate sequences of various shapes conditioned on the provided initial frames. Previous approaches achieve conditional human motion generation on video~\cite{srivastava2015unsupervised} and skeletons~\cite{cai2017deep}, but not on 3D shape sequences. To illustrate the effectiveness of our method, we take Dyna~\cite{pons2015dyna} as an example. In this collection of datasets, there are female/male models of different subjects and actions. All meshes in different datasets have the same number of vertices and share connectivity, so we train our model on a mixture of those datasets. In testing, we feed $u(=2)$ bootstrap models with a certain body mass index (BMI)/gender/motion as input, and get the following $n(=16)$ sequences as output. We show our results in Fig.~\ref{fig:conditional}. The observation is that our method can generate human shapes in different subjects and gender. Furthermore, even if the first frame is the same, the network can produce different action sequences according to the second frame.

	\subsection{Sequence Completion}  \label{sec:completion}
	
	We now consider another important application namely sequence completion, which produces in-between shapes given two endpoint frames.

	\textbf{Completion based on key frames.}
	
	Completing a mesh sequence based on given anchors is an important application in animation. In our approach, we clip the target sequence by keyframes. For each segment, we run our bidirectional network by treating two keyframes as endpoints. Once the forward and backward sequences converge at a model, we stitch them to form a whole sequence. Since the computation is identical for each segment, for illustration we show an example of completing one segment constrained on two key frames. Fig~\ref{fig:completion} shows an example on the Dyna Dataset~\cite{pons2015dyna}. (f)(b)(h) are all interpolation-based. Those methods generate shapes along the shortest path between them, which are almost still because of high similarity between the first and last models.
	
	\textbf{Generating novel sequence.}
	Previous interpolation-based methods usually adopt a deterministic strategy to complete sequences, and thus result in a monotonous sequence. Our work, however, is able to produce diversified sequence completion results. By assigning a random vector to the LSTM state, the network generates different sequences as shown in Fig.~\ref{fig:completion}. In the real world, there are often more than one possible motions between two static poses and our model can therefore better describe such characteristics in human motion than other generation methods. \\
	To test alternative completion strategies, we also implement an optimization+unidirection~\cite{cai2017deep} strategy. Given the source model $X_0$ and $X_n$,  we first find the optimal LSTM initial state $\hat{s_0} = \underset{s_0}{\arg\min}||\hat{X_n}-X_n|| $, where	
	\begin{equation}\label{equa:s}
	\left\{
	\begin{aligned}
	(\hat{X_t},s_t)&=G(X_0,s_0)                           & i=t \\
	(\hat{X}_{t+1},s_{t+1})&=G(\hat{X_t},s_t)  & i > t
	\end{aligned}
	\right.
	\end{equation}
	After solving the optimization problem with~\cite{hansen2001completely}, we then compute $\{\hat{X_t}\}$ through Eq.~\ref{equa:s}. The result is shown in Fig.~\ref{fig:completion} (e).
	Compared to interpolation strategies, the optimization+unidirection algorithm can achieve more realistic morphing, but it does not provide diverse possible choices as our BD-approach.

	\subsection{Implementation Details}\label{sec:Implementation}
	We use Tensorflow as the framework of our implementation. Experiments are performed on a PC with an Intel Core i7-2600 CPU and an NVIDIA Tesla K40c GPU. We use Adam optimizer~\cite{kingma2014adam} to update weights, with default Adam parameters $\beta _1=0.9,\beta _2=0.999$ as in \cite{kingma2014adam}. For each dataset, we randomly exclude a subsequence, which takes up 20\% of the dataset, as a a test set. A training process takes 7000 iterations, lasting for around 8 hours. In each iteration, we generate 8 sequences, each of them containing 32 shapes. For the dataset where motion is slow~\cite{pons2015dyna}, we sample every other model in sequences. For all experiments, the LSTM $cell$ has 3 layers and 128 hidden dimensions, and we set initial states as $s^{f}_0=-s^{b}_0=[0.1]^{128}$. The mesh convolution module $Conv$ is composed of 3 layers with $tanh$ as the activation function. Transpose convolutions $tCnv$ mirrors $Conv$ and shares the same weights.

	\section{Conclusion}
	In this paper, we propose the first deep architecture to generate mesh animation sequences, which can not only predict future frames given initial frames, but also complete mesh sequences based on key frames and generate sequences conditioned on given shapes.
	Extensive qualitative and quantitative evaluation demonstrates that our method achieves state-of-the-art generation results, and our completion strategy is also able to produce diverse realistic results.
	{\small
		\bibliography{main}

\begin{thebibliography}{}

\bibitem[\protect\citeauthoryear{Bogo \bgroup et al\mbox.\egroup
  }{2014}]{Bogo2014}
Bogo, F.; Romero, J.; Loper, M.; and Black, M.~J.
\newblock 2014.
\newblock {FAUST}: Dataset and evaluation for {3D} mesh registration.
\newblock In {\em Proceedings IEEE Conf. on Computer Vision and Pattern
  Recognition (CVPR)},  3794 --3801.

\bibitem[\protect\citeauthoryear{Bowman \bgroup et al\mbox.\egroup
  }{2015}]{bowman2015generating}
Bowman, S.~R.; Vilnis, L.; Vinyals, O.; Dai, A.~M.; Jozefowicz, R.; and Bengio,
  S.
\newblock 2015.
\newblock Generating sentences from a continuous space.
\newblock {\em arXiv preprint arXiv:1511.06349}.

\bibitem[\protect\citeauthoryear{Bronstein \bgroup et al\mbox.\egroup
  }{2017}]{bronstein2017geometric}
Bronstein, M.~M.; Bruna, J.; LeCun, Y.; Szlam, A.; and Vandergheynst, P.
\newblock 2017.
\newblock Geometric deep learning: going beyond euclidean data.
\newblock {\em IEEE Signal Processing Magazine} 34(4):18--42.

\bibitem[\protect\citeauthoryear{B{\"u}tepage \bgroup et al\mbox.\egroup
  }{2017}]{butepage2017deep}
B{\"u}tepage, J.; Black, M.~J.; Kragic, D.; and Kjellstr{\"o}m, H.
\newblock 2017.
\newblock Deep representation learning for human motion prediction and
  classification.
\newblock In {\em IEEE Conference on Computer Vision and Pattern Recognition
  (CVPR)},  2017.

\bibitem[\protect\citeauthoryear{Cai \bgroup et al\mbox.\egroup
  }{2017}]{cai2017deep}
Cai, H.; Bai, C.; Tai, Y.-W.; and Tang, C.-K.
\newblock 2017.
\newblock Deep video generation, prediction and completion of human action
  sequences.
\newblock {\em arXiv:1711.08682}.

\bibitem[\protect\citeauthoryear{Cho \bgroup et al\mbox.\egroup
  }{2014}]{cho2014learning}
Cho, K.; Van~Merri{\"e}nboer, B.; Gulcehre, C.; Bahdanau, D.; Bougares, F.;
  Schwenk, H.; and Bengio, Y.
\newblock 2014.
\newblock Learning phrase representations using {RNN} encoder-decoder for
  statistical machine translation.
\newblock {\em arXiv preprint arXiv:1406.1078}.

\bibitem[\protect\citeauthoryear{Chung \bgroup et al\mbox.\egroup
  }{2015}]{chung2015recurrent}
Chung, J.; Kastner, K.; Dinh, L.; Goel, K.; Courville, A.~C.; and Bengio, Y.
\newblock 2015.
\newblock A recurrent latent variable model for sequential data.
\newblock In {\em NIPS},  2980--2988.

\bibitem[\protect\citeauthoryear{Dou \bgroup et al\mbox.\egroup
  }{2016}]{Dou2016}
Dou, M.; Khamis, S.; Degtyarev, Y.; Davidson, P.; Fanello, S.~R.; Kowdle, A.;
  Escolano, S.~O.; Rhemann, C.; Kim, D.; Taylor, J.; Kohli, P.; Tankovich, V.;
  and Izadi, S.
\newblock 2016.
\newblock Fusion4d: Real-time performance capture of challenging scenes.
\newblock {\em ACM Trans. Graph.} 35(4):114:1--114:13.

\bibitem[\protect\citeauthoryear{Duvenaud \bgroup et al\mbox.\egroup
  }{2015}]{duvenaud2015convolutional}
Duvenaud, D.~K.; Maclaurin, D.; Iparraguirre, J.; Bombarell, R.; Hirzel, T.;
  Aspuru-Guzik, A.; and Adams, R.~P.
\newblock 2015.
\newblock Convolutional networks on graphs for learning molecular fingerprints.
\newblock In {\em NIPS},  2224--2232.

\bibitem[\protect\citeauthoryear{Fragkiadaki \bgroup et al\mbox.\egroup
  }{2015}]{fragkiadaki2015recurrent}
Fragkiadaki, K.; Levine, S.; Felsen, P.; and Malik, J.
\newblock 2015.
\newblock Recurrent network models for human dynamics.
\newblock In {\em IEEE International Conference on Computer Vision (ICCV)},
  4346--4354.

\bibitem[\protect\citeauthoryear{Gao \bgroup et al\mbox.\egroup
  }{2017a}]{gao2017data}
Gao, L.; Chen, S.-Y.; Lai, Y.-K.; and Xia, S.
\newblock 2017a.
\newblock Data-driven shape interpolation and morphing editing.
\newblock {\em Computer Graphics Forum} 36(8):19--31.

\bibitem[\protect\citeauthoryear{Gao \bgroup et al\mbox.\egroup
  }{2017b}]{gao2017sparse}
Gao, L.; Lai, Y.-K.; Yang, J.; Zhang, L.-X.; Kobbelt, L.; and Xia, S.
\newblock 2017b.
\newblock Sparse data driven mesh deformation.
\newblock {\em arXiv preprint arXiv:1709.01250}.

\bibitem[\protect\citeauthoryear{Goodfellow \bgroup et al\mbox.\egroup
  }{2014}]{goodfellow2014generative}
Goodfellow, I.; Pouget-Abadie, J.; Mirza, M.; Xu, B.; Warde-Farley, D.; Ozair,
  S.; Courville, A.; and Bengio, Y.
\newblock 2014.
\newblock Generative adversarial nets.
\newblock In {\em NIPS},  2672--2680.

\bibitem[\protect\citeauthoryear{Gregor \bgroup et al\mbox.\egroup
  }{2015}]{gregor2015draw}
Gregor, K.; Danihelka, I.; Graves, A.; Rezende, D.~J.; and Wierstra, D.
\newblock 2015.
\newblock Draw: A recurrent neural network for image generation.
\newblock {\em arXiv preprint arXiv:1502.04623}.

\bibitem[\protect\citeauthoryear{Hansen and
  Ostermeier}{2001}]{hansen2001completely}
Hansen, N., and Ostermeier, A.
\newblock 2001.
\newblock Completely derandomized self-adaptation in evolution strategies.
\newblock {\em Evolutionary computation} 9(2):159--195.

\bibitem[\protect\citeauthoryear{Hochreiter and
  Schmidhuber}{1997}]{hochreiter1997long}
Hochreiter, S., and Schmidhuber, J.
\newblock 1997.
\newblock Long short-term memory.
\newblock {\em Neural computation} 9(8):1735--1780.

\bibitem[\protect\citeauthoryear{Huang, Kalogerakis, and
  Marlin}{2015}]{huang2015analysis}
Huang, H.; Kalogerakis, E.; and Marlin, B.
\newblock 2015.
\newblock Analysis and synthesis of 3d shape families via deep-learned
  generative models of surfaces.
\newblock {\em Computer Graphics Forum} 34(5):25--38.

\bibitem[\protect\citeauthoryear{Huber, Perl, and
  Rumpf}{2017}]{huber2017smooth}
Huber, P.; Perl, R.; and Rumpf, M.
\newblock 2017.
\newblock Smooth interpolation of key frames in a riemannian shell space.
\newblock {\em Computer Aided Geometric Design} 52:313--328.

\bibitem[\protect\citeauthoryear{Kalogerakis \bgroup et al\mbox.\egroup
  }{2017}]{kalogerakis20173d}
Kalogerakis, E.; Averkiou, M.; Maji, S.; and Chaudhuri, S.
\newblock 2017.
\newblock {3D} shape segmentation with projective convolutional networks.
\newblock In {\em IEEE Conference on Computer Vision and Pattern Recognition
  (CVPR)}.

\bibitem[\protect\citeauthoryear{Kingma and Ba}{2014}]{kingma2014adam}
Kingma, D.~P., and Ba, J.
\newblock 2014.
\newblock Adam: A method for stochastic optimization.
\newblock {\em arXiv preprint arXiv:1412.6980}.

\bibitem[\protect\citeauthoryear{Levi and Gotsman}{2015}]{Levi}
Levi, Z., and Gotsman, C.
\newblock 2015.
\newblock Smooth rotation enhanced as-rigid-as-possible mesh animation.
\newblock {\em IEEE Trans. Vis. Comp. Graph.} 21(2):264--277.

\bibitem[\protect\citeauthoryear{Li \bgroup et al\mbox.\egroup
  }{2017}]{li2017auto}
Li, Z.; Zhou, Y.; Xiao, S.; He, C.; and Li, H.
\newblock 2017.
\newblock Auto-conditioned lstm network for extended complex human motion
  synthesis.
\newblock {\em arXiv preprint arXiv:1707.05363}.

\bibitem[\protect\citeauthoryear{Lotter, Kreiman, and
  Cox}{2016}]{lotter2016deep}
Lotter, W.; Kreiman, G.; and Cox, D.
\newblock 2016.
\newblock Deep predictive coding networks for video prediction and unsupervised
  learning.
\newblock {\em arXiv:1605.08104}.

\bibitem[\protect\citeauthoryear{Lyu \bgroup et al\mbox.\egroup
  }{2015}]{lyu2015modelling}
Lyu, Q.; Wu, Z.; Zhu, J.; and Meng, H.
\newblock 2015.
\newblock Modelling high-dimensional sequences with {LSTM-RTRBM}: Application
  to polyphonic music generation.
\newblock In {\em IJCAI},  4138--4139.

\bibitem[\protect\citeauthoryear{Marchi \bgroup et al\mbox.\egroup
  }{2014}]{marchi2014multi}
Marchi, E.; Ferroni, G.; Eyben, F.; Gabrielli, L.; Squartini, S.; and Schuller,
  B.
\newblock 2014.
\newblock Multi-resolution linear prediction based features for audio onset
  detection with bidirectional lstm neural networks.
\newblock In {\em IEEE International Conference on Acoustics, Speech and Signal
  Processing (ICASSP)},  2164--2168.

\bibitem[\protect\citeauthoryear{Martinez, Black, and
  Romero}{2017}]{martinez2017human}
Martinez, J.; Black, M.~J.; and Romero, J.
\newblock 2017.
\newblock On human motion prediction using recurrent neural networks.
\newblock In {\em IEEE Conference on Computer Vision and Pattern Recognition
  (CVPR)},  4674--4683.

\bibitem[\protect\citeauthoryear{Mathieu, Couprie, and
  LeCun}{2015}]{mathieu2015deep}
Mathieu, M.; Couprie, C.; and LeCun, Y.
\newblock 2015.
\newblock Deep multi-scale video prediction beyond mean square error.
\newblock {\em arXiv preprint arXiv:1511.05440}.

\bibitem[\protect\citeauthoryear{Melamud, Goldberger, and
  Dagan}{2016}]{melamud2016context2vec}
Melamud, O.; Goldberger, J.; and Dagan, I.
\newblock 2016.
\newblock Context2vec: Learning generic context embedding with bidirectional
  {LSTM}.
\newblock In {\em {SIGNLL} Conference on Computational Natural Language
  Learning},  51--61.

\bibitem[\protect\citeauthoryear{Mikolov \bgroup et al\mbox.\egroup
  }{2011}]{mikolov2011extensions}
Mikolov, T.; Kombrink, S.; Burget, L.; {\v{C}}ernock{\`y}, J.; and Khudanpur,
  S.
\newblock 2011.
\newblock Extensions of recurrent neural network language model.
\newblock In {\em IEEE International Conference on Acoustics, Speech and Signal
  Processing (ICASSP)},  5528--5531.

\bibitem[\protect\citeauthoryear{Oh \bgroup et al\mbox.\egroup
  }{2015}]{oh2015action}
Oh, J.; Guo, X.; Lee, H.; Lewis, R.~L.; and Singh, S.
\newblock 2015.
\newblock Action-conditional video prediction using deep networks in atari
  games.
\newblock In {\em NIPS},  2863--2871.

\bibitem[\protect\citeauthoryear{Pons-Moll \bgroup et al\mbox.\egroup
  }{2015}]{pons2015dyna}
Pons-Moll, G.; Romero, J.; Mahmood, N.; and Black, M.~J.
\newblock 2015.
\newblock Dyna: A model of dynamic human shape in motion.
\newblock {\em ACM Transactions on Graphics (TOG)} 34(4):120.

\bibitem[\protect\citeauthoryear{Riegler, Ulusoy, and
  Geiger}{2017}]{riegler2017octnet}
Riegler, G.; Ulusoy, A.~O.; and Geiger, A.
\newblock 2017.
\newblock Octnet: Learning deep {3D} representations at high resolutions.
\newblock In {\em IEEE Conference on Computer Vision and Pattern Recognition
  (CVPR)}, volume~3.

\bibitem[\protect\citeauthoryear{Sidi \bgroup et al\mbox.\egroup
  }{2011}]{sidi2011unsupervised}
Sidi, O.; van Kaick, O.; Kleiman, Y.; Zhang, H.; and Cohen-Or, D.
\newblock 2011.
\newblock Unsupervised co-segmentation of a set of shapes via descriptor-space
  spectral clustering.
\newblock {\em ACM Transactions on Graphics (TOG)} 30(6).

\bibitem[\protect\citeauthoryear{Srivastava, Mansimov, and
  Salakhudinov}{2015}]{srivastava2015unsupervised}
Srivastava, N.; Mansimov, E.; and Salakhudinov, R.
\newblock 2015.
\newblock Unsupervised learning of video representations using lstms.
\newblock In {\em International Conference on Machine Learning},  843--852.

\bibitem[\protect\citeauthoryear{Stoll \bgroup et al\mbox.\egroup
  }{2010}]{Stoll2010}
Stoll, C.; Gall, J.; de~Aguiar, E.; Thrun, S.; and Theobalt, C.
\newblock 2010.
\newblock Video-based reconstruction of animatable human characters.
\newblock {\em ACM Trans. Graph.} 29(6):139:1--139:10.

\bibitem[\protect\citeauthoryear{Su \bgroup et al\mbox.\egroup
  }{2015}]{su2015multi}
Su, H.; Maji, S.; Kalogerakis, E.; and Learned-Miller, E.
\newblock 2015.
\newblock Multi-view convolutional neural networks for {3D} shape recognition.
\newblock In {\em IEEE International Conference on Computer Vision},  945--953.

\bibitem[\protect\citeauthoryear{Sumner and
  Popovi{\'c}}{2004}]{sumner2004deformation}
Sumner, R.~W., and Popovi{\'c}, J.
\newblock 2004.
\newblock Deformation transfer for triangle meshes.
\newblock {\em ACM Transactions on Graphics (TOG)} 23(3):399--405.

\bibitem[\protect\citeauthoryear{Tan \bgroup et al\mbox.\egroup
  }{2018a}]{tan2017variational}
Tan, Q.; Gao, L.; Lai, Y.-K.; and Xia, S.
\newblock 2018a.
\newblock Variational autoencoders for deforming 3d mesh models.
\newblock In {\em IEEE Conference on Computer Vision and Pattern Recognition
  (CVPR)}.

\bibitem[\protect\citeauthoryear{Tan \bgroup et al\mbox.\egroup
  }{2018b}]{tan2017mesh}
Tan, Q.; Gao, L.; Lai, Y.-K.; Yang, J.; and Xia, S.
\newblock 2018b.
\newblock Mesh-based autoencoders for localized deformation component analysis.
\newblock In {\em AAAI}.

\bibitem[\protect\citeauthoryear{Vinyals \bgroup et al\mbox.\egroup
  }{2015}]{vinyals2015show}
Vinyals, O.; Toshev, A.; Bengio, S.; and Erhan, D.
\newblock 2015.
\newblock Show and tell: A neural image caption generator.
\newblock In {\em IEEE Conference on Computer Vision and Pattern Recognition
  (CVPR)},  3156--3164.

\bibitem[\protect\citeauthoryear{Vlasic \bgroup et al\mbox.\egroup
  }{2008}]{vlasic2008articulated}
Vlasic, D.; Baran, I.; Matusik, W.; and Popovi{\'c}, J.
\newblock 2008.
\newblock Articulated mesh animation from multi-view silhouettes.
\newblock {\em ACM Transactions on Graphics (TOG)} 27(3):97.

\bibitem[\protect\citeauthoryear{Vondrick, Pirsiavash, and
  Torralba}{2016}]{vondrick2016generating}
Vondrick, C.; Pirsiavash, H.; and Torralba, A.
\newblock 2016.
\newblock Generating videos with scene dynamics.
\newblock In {\em NIPS},  613--621.

\bibitem[\protect\citeauthoryear{Walker \bgroup et al\mbox.\egroup
  }{2017}]{walker2017pose}
Walker, J.; Marino, K.; Gupta, A.; and Hebert, M.
\newblock 2017.
\newblock The pose knows: Video forecasting by generating pose futures.
\newblock In {\em IEEE International Conference on Computer Vision (ICCV)},
  3352--3361.

\bibitem[\protect\citeauthoryear{Wu \bgroup et al\mbox.\egroup
  }{2016}]{wu2016learning}
Wu, J.; Zhang, C.; Xue, T.; Freeman, B.; and Tenenbaum, J.
\newblock 2016.
\newblock Learning a probabilistic latent space of object shapes via {3D}
  generative-adversarial modeling.
\newblock In {\em NIPS},  82--90.

\bibitem[\protect\citeauthoryear{Yu \bgroup et al\mbox.\egroup
  }{2017}]{yu2017seqgan}
Yu, L.; Zhang, W.; Wang, J.; and Yu, Y.
\newblock 2017.
\newblock {SeqGAN}: Sequence generative adversarial nets with policy gradient.
\newblock In {\em AAAI},  2852--2858.

\end{thebibliography}
		\bibliographystyle{aaai}
	}
\end{document}